# Energy-Efficient Hybrid Stochastic-Binary Neural Networks for Near-Sensor Computing

Vincent T. Lee[†], Armin Alaghi[†], John P. Hayes[*], Visvesh Sathe[‡], Luis Ceze[†]
[†]Department of Computer Science and Engineering, University of Washington, Seattle, WA, 98198
[*]Department of Electrical Engineering and Computer Science, University of Michigan, Ann Arbor, MI, 48109
[‡]Department of Electrical Engineering, University of Washington, Seattle, WA, 98195
{vlee2, armin}@cs.washington.edu, jhayes@eecs.umich.edu, sathe@uw.edu, luisceze@cs.washington.edu

*Abstract*— **Recent advances in neural networks (NNs) exhibit unprecedented success at transforming large, unstructured data streams into compact higher-level semantic information for tasks such as handwriting recognition, image classification, and speech recognition. Ideally, systems would employ near-sensor computation to execute these tasks at sensor endpoints to maximize data reduction and minimize data movement. However, near-sensor computing presents its own set of challenges such as operating power constraints, energy budgets, and communication bandwidth capacities. In this paper, we propose a stochastic-binary hybrid design which splits the computation between the stochastic and binary domains for near-sensor NN applications. In addition, our design uses a new stochastic adder and multiplier that are significantly more accurate than existing adders and multipliers. We also show that retraining the binary portion of the NN computation can compensate for precision losses introduced by shorter stochastic bit-streams, allowing faster run times at minimal accuracy losses. Our evaluation shows that our hybrid stochastic-binary design can achieve 9.8× energy efficiency savings, and application-level accuracies within 0.05% compared to conventional all-binary designs.**

## I. INTRODUCTION

Sensors and actuators are critical for enabling electronic circuits to interact with the physical world. Information acquired from sensors has become essential to applications from home automation to medical implants to environmental surveillance. It is predicted that the world soon will have an average of 1,000 sensors per person [8][11] which translates to a huge amount of raw data acquisition. The sheer volume of unstructured sensor data threatens to overwhelm storage and network communication capacities, which are increasingly limited by aggressive power and energy budgets.

To reduce the storage and communication demands of raw sensor data, *near-sensor computing* has recently emerged as a design space for reducing these overheads [20]. Near-sensor computing proposes offloading portions of the application to computing units or accelerators co-located with the sensing device. The key insight is that by offloading certain portions of computation such as image feature extraction (of an image-processing pipeline) to sensor end points, higher level semantic information can be transmitted in place of larger unstructured data streams. Of particular interest are neural networks (NNs) which are a widely used class of algorithms for processing raw unstructured data. NNs excel at reasoning about raw data streams in applications such as object detection, handwriting recognition, and speech processing. Recent work by Du et al. [12] shows how a near-sensor NN accelerator can dramatically reduce the energy costs of the system.

This paper presents a near-sensor stochastic-binary NN design which combines stochastic computing (SC) with conventional "binary" processing and sensor data acquisition to improve energy efficiency and power consumption. SC is a re-emerging computation technique that performs computation on unary bit-streams representing probabilities [14]. SC circuits are often cheaper than binary arithmetic circuits [25]. For instance, multiplication in SC can be implemented by a single AND gate. The primary tradeoff for SC's simplicity is increased computation time, which leads to higher energy consumption for higher precision calculations [2][22]. However, for applications that can tolerate reduced precision, SC can achieve compelling power and energy efficiency gains. Finally, stochastic circuits are smaller in size and more error tolerant, making them suitable for tiny sensors operating in harsh environments [3][13].

Stochastic NNs have been extensively studied in the prior literature [7][9][15]. However, past work proposes fully stochastic designs that have number lengths exceeding 1,000 clock cycles [7][15], which leads to higher energy consumption. In addition, errors introduced by multiple levels of SC circuits compound as more levels are executed [22]. In this paper, we present a stochastic-binary hybrid NN system that exploits the benefits of SC, while mitigating many of its drawbacks. We only employ SC in the first layer of an NN, so it operates directly on the sensor data thereby avoiding the issue of compounding errors over multiple layers. We employ a new, significantly more accurate SC adder and a deterministic number generation scheme to further reduce energy consumption. Finally, we compare our design's accuracy to that of existing SC designs, and show our design has better energy efficiency than competing binary implementations.

Our contributions are as follows:

1. A novel stochastic adder for convolutional NNs which increases speed and/or accuracy, leading to a reduced energy cost compared to previous SC NN designs.
2. A hybrid stochastic-binary NN design which combines signal acquisition and SC in the first NN layer, and uses binary for the remaining layers to avoid compounding accuracy losses.
3. Showing that retraining these remaining NN layers can compensate for precision loss introduced by SC.

The rest of the paper is organized as follows. Section II provides background on SC and NNs. Section III introduces the new stochastic adder design. Section IV presents our hybrid NN design, and results are discussed in Sections V and VI.

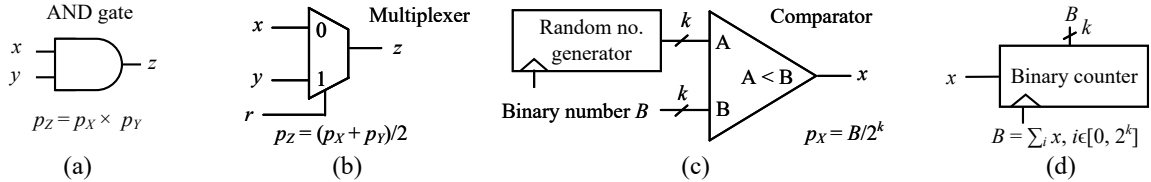

Fig. 1. Unipolar stochastic arithmetic primitives: (a) multiplier, (b) scaled adder, (c) comparator-based stochastic number generator, and (d) stochastic-to-binary converter implemented as a binary counter.

## II. BACKGROUND

This section briefly reviews the relevant concepts of stochastic computing and neural networks.

### A. Stochastic Computing

Stochastic computing is an alternative method of computing first proposed in the 1960s [14]. In SC, numbers are encoded as bit-streams that are interpreted as probabilities. For instance, the bit-stream $X = 001011\ldots$ denotes a stochastic number (SN) with value $p_X = 0.5$ because the probability of seeing a 1 at a randomly chosen position of $X$ is 0.5. This interpretation allows arithmetic functions to be implemented via simple logic gates. For instance, the AND gate in Fig. 1a performs multiplication on uncorrelated inputs. The SC probability $p_X$ or *unipolar* range [0, 1], does not include negative numbers, which are usually needed for NNs. As a result, NNs often use *bipolar* numbers, where the value of $X$ is interpreted as $2p_X - 1$, and therefore has range [−1, 1]. The precision of SC is mainly determined by the length $N$ of the bit-stream. A bit-stream of length $N$ encodes a number at $\log_2 N$ bits of precision. For example, a unipolar bit-stream of length 16 can encode the range [0, 15] which is equivalent to the range of a binary number with $\log_2 16 = 4$ bits of precision.

In this work, we use four SC primitives: adders, multipliers, stochastic number generators (SNGs), and stochastic-to-digital converters (Fig. 1). These components operate on unipolar numbers; they may implement a different function when interpreted in the bipolar domain. To perform conventional stochastic addition, two bit-streams $X$ and $Y$ are applied to the data inputs of a multiplexer with the select bit driven by a bit-stream $R$ of unipolar value $p_R = 0.5$ (Fig. 1b). The output bit-stream encodes $p_Z = 0.5(p_X + p_Y)$. Notice the scale factor of 0.5, a necessary feature of SC, keeps the probability in the unit interval [0, 1]. When compounded over many additions, the scale factor can lead to severe loss of precision. Similar precision losses also occur with SC multiplication, which is realized with an AND gate (Fig. 1a) since $p_Z = p_X \times p_Y$. One way to improve the quality of a function is to increase the length of the input bit-streams. However, since each bit of additional precision requires a doubling of bit-stream length this quickly leads to excessive run times. As a result, researchers have proposed alternative designs that approximate the add operation. One example is to use an OR gate as an adder, which only works accurately if both inputs are close to zero [21]. Hence, all existing SC adder designs need additional uncorrelated random number sources and/or have limited accuracy. The need for extra random number sources becomes severe when many numbers are to be added. Ideally, we would like an adder that operates accurately on many inputs in short periods of time, without requiring additional uncorrelated number sources.

Binary-to-stochastic converters, which are commonly referred to as stochastic number generators (SNGs), and stochastic-to-binary converters are SC primitives that allow conversion between the binary and stochastic domains. An SNG comprises a comparator and a random number generator (Fig. 1c). For a given number $p_X$, the SNG will produce a 1 with that probability if the random number is less than $p_X$. Converting analog signals to the stochastic domain can be achieved by replacing the SNG comparator with an analog one. In this paper, we use an analog-to-stochastic converter to convert the sensor data directly to stochastic encodings, without the need for analog-to-digital converters (ADCs). We also use a set of SNGs to generate the NN weights.

The choice of SNG configuration affects the accuracy and consequently the energy consumption of the SC circuit. Table 1 shows the mean square error (MSE) of a 4-bit and 8-bit SC multiplier for the following SNG schemes: (i) using the same linear feedback shift register (LFSR) for both inputs, (ii) using a separate LFSR for each input, (iii) using low-discrepancy sequences [4], and (iv) using a ramp-compare analog-to-stochastic converter [13] for one input, and a low-discrepancy sequence for the other. For this work, we employ the last number generation scheme as it provides the best accuracy. The MSEs are calculated by exhaustively testing the multipliers for every possible input value.

To convert from stochastic to binary, we simply count the 1s in the bit-stream by using a binary counter (Fig. 1d). In our work, we use asynchronous counters because they allow us to clock the SC part of the circuit faster. It is sufficient to apply a new input to an asynchronous counter, even if the previous inputs have not propagated through the counter. The delay of a synchronous counter, on the other hand, is relatively large, so it cannot keep up with the speed of the SC circuit feeding it. Unlike the asynchronous counter, a synchronous counter fails if the next input arrives before the previous input is propagated.

### B. Neural Networks

NNs come in a wide range of network topologies, and generally consist of an input layer, an output layer, and a number of hidden layers in between [24]. A layer is composed of neurons, each of which has a set of inputs, an output, and an activation function $f(x)$, e.g., a rectified linear unit. Each neuron is connected to neurons in the previous layer; a connection is defined by a weight that is multiplied by the previous neuron's output. These values are summed with other connections' outputs and passed to an activation function. For instance, given a neuron $y$ that is connected to $k$ neurons in the previous layer with output values $\vec{x} = \{x_0, x_1, …, x_{k-1}\}$ and connection weights $\vec{w} = \{w_0, w_1, …, w_{k-1}\}$ respectively, the output of neuron $y$ is defined as $y_{out} = f(\sum_{i=0}^{k-1} x_i w_i)$.

Neuron connection topologies can either be fully connected or locally connected to the previous layer. In fully connected layers, each neuron is connected to every neuron of the previous layer. In the locally connected case, neurons are connected to a subset of neurons in the previous layer. Locally connected layers are often referred to as convolutional layers because their connections from the previous layer take the form of a window. The resulting operation is mathematically equivalent to a convolution where the convolutional kernel is simply a matrix of the connection weights. Finally, NNs also may have max pooling layers, which are locally connected layers that subsample a window in the previous layer and output the maximum value.

To determine the weights for each layer, NNs are trained over an input training set using backpropagation [24]. This is a technique that iterates over the training dataset and gradually adjusts the weights based on the gradient of the error in the NN's output function. The error metric varies across applications but a commonly used one for NN classification is the cross-entropy loss. One iteration over the entire training set is known as an epoch. Training is often supplemented by dropout which is a training technique that randomly removes connections during the training process at certain layers to prevent overfitting. Once the training process converges to a set of weights, a test set is used to evaluate the quality of the NN model. The quality metric varies across applications but a commonly used metric is classification accuracy based on the outputs of the NN model.

Using SC for NNs has a well-established history [7][17] dating back to the 1990s. Recent work proposes fully stochastic NN designs using FPGA fabrics and full custom ASICs [16]. Similarly, Ardakani et al. [6] propose an SC NN for digit recognition which outperforms binary designs by using shorter bit-streams (down to length 16). To the best of our knowledge, this is the only SC NN design that outperforms, albeit marginally, its binary counterpart in terms of energy efficiency. However, unlike our approach, prior SC work uses older, fully connected NN topologies with only two hidden layers which are smaller and less accurate than current state-of-the art NN topologies like LeNet-5 (used in our evaluation). Finally, fully stochastic NNs need longer bit-streams ($N$ = 256 to 1024) to achieve reasonable accuracy. In contrast, our work does not execute the entire NN in the stochastic domain. Instead, we execute the first layer using SC, then allow higher precision binary units to finish the NN calculation.

III. STOCHASTIC ADDER DESIGN

Unlike the basic stochastic multiplier, the conventional stochastic add operation has undesirable properties such as the enforced scaling factor and an extra bit-stream. Furthermore, the discarding of some bits of each number (through multiplexing) leads to accuracy loss, which compounds with multiple additions.

Table 1. MSE of stochastic multiplier for different RNG methods (lower is better)

| Number generation scheme | 8-Bit Prec. | 4-Bit Prec. |
|---|---|---|
| One LFSR + shifted version | 2.78×10⁻³ | 2.99×10⁻³ |
| Two LFSRs | 2.57×10⁻⁴ | 1.60×10⁻³ |
| Low-discrepancy sequences [4] | 1.28×10⁻⁵ | 1.01×10⁻³ |
| Ramp-compare [13] + [4] | **8.66×10⁻⁶** | **7.21×10⁻⁴** |

Table 2. MSE of stochastic addition for different SNG methods (lower is better)

| Implementation | | 8-Bit Prec. | 4-Bit Prec. |
|---|---|---|---|
| Old adder (Fig. 1b) | Random + LFSR | 3.24×10⁻⁴ | 5.55×10⁻³ |
| | Random + TFF | 5.49×10⁻⁴ | 5.49×10⁻³ |
| | LFSR + TFF | 1.06×10⁻⁴ | 2.66×10⁻³ |
| New adder (Fig. 2b) | | **1.91×10⁻⁶** | **4.88×10⁻⁴** |

We now propose a new stochastic adder that is more accurate and does not require additional random inputs. But first we introduce a simple circuit that implements the SC function $p_C = p_A/2$. A rudimentary implementation is to use the multiplier of Fig. 1a where we assign $A$ to one input, and a randomly generated bit-stream $B$ of value 1/2 to the other. Note that for the multiplication to work accurately, $B$ has to be uncorrelated to $A$. Fig. 2a shows another implementation of the same function, in which a bit-stream $B$ with value 1/2 is generated from the bit-stream of $A$ without requiring an additional input. A toggle flip-flop (TFF), which switches its output between 0 and 1 when its input is 1, is used for this purpose. The area cost of a TFF is no more than a random number generator that is required for generating 1/2. More importantly, the bit-stream generated by the TFF is always uncorrelated with its input bit-stream. This means that there are no constraints on the auto-correlation of the input bit-stream, unlike common sequential SC circuits that do not function as intended if the input is auto-correlated [7].

Fig. 2b shows our proposed TFF-based adder. At each clock cycle, if the values at $X$ and $Y$ are equal, they propagate to the output. Otherwise, the state of the TFF is output and the TFF is toggled. Suppose the adder operates on two bit-streams of length 20. Recall for adds, there is a 0.5 normalization constant, so the expected result is $Z = 0.5(1/2 + 4/5) = 13/20$ computed as follows:

$X$ = 0110 0011 0101 0111 1000 (1/2)
$Y$ = 1011 1111 0101 0111 1111 (4/5)
$Z$ = 0110 1011 0101 0111 1101 (13/20)

The result of the adder is always accurate if the bit-stream length $N$ is sufficient to represent it. Otherwise, the output will be rounded off to the nearest representable number. The direction of rounding depends on the initial state $S_0$ of the TFF.

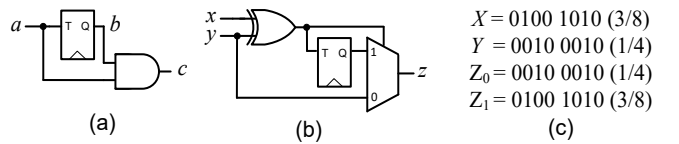

Fig. 2. (a) Stochastic circuit with $p_C = p_A/2$, (b) proposed TFF-based stochastic adder with $p_Z = (p_X + p_Y)/2$, and (c) example of its operation with two different initial states.

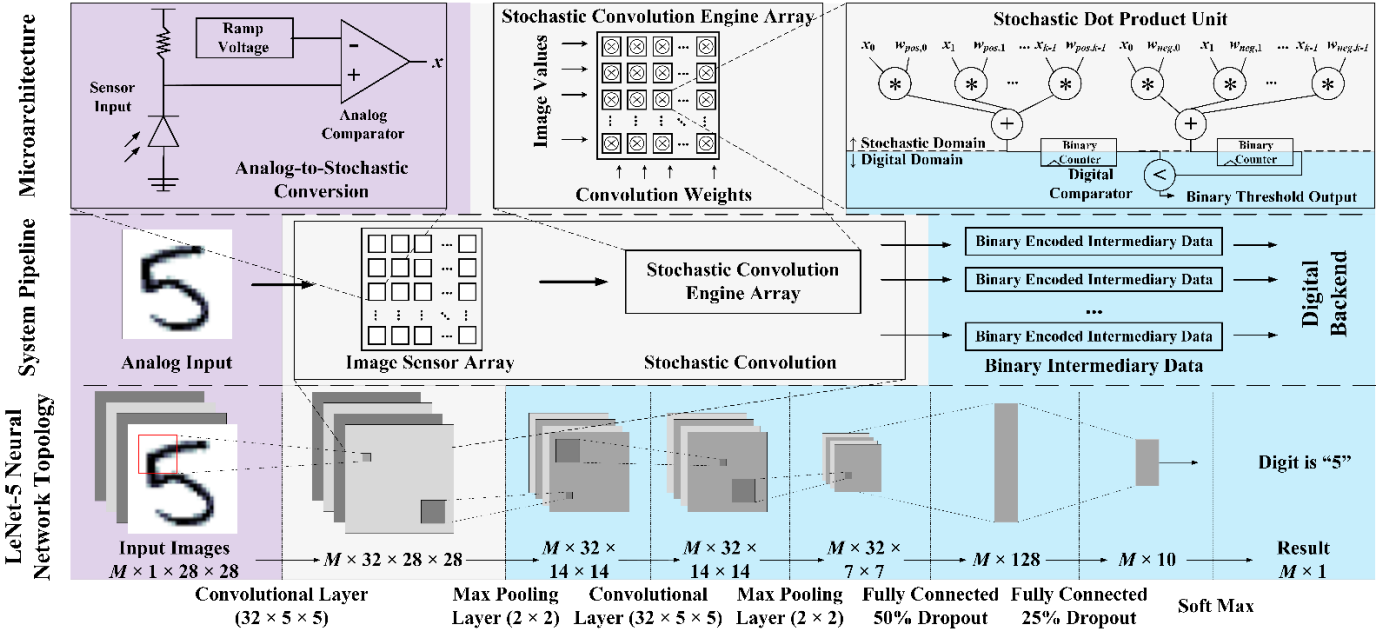

Fig. 3. System diagram of our proposed near-sensor stochastic NN. Bottom – LeNet-5 NN topology. Middle – system pipeline. Top – microarchitecture. Purple, grey, and blue regions denote analog, stochastic, and binary domains, respectively.

If $S_0 = 0$, as in the example above, the result will be rounded to the smaller of the two neighboring numbers. Fig. 2c shows how $S_0$ affects the result. $Z_0$ and $Z_1$ are the outputs of the circuit with $S_0 = 0$ and 1, respectively. The expected result is $Z = 0.5(3/8 + 1/4) = 5/16$. Since $N = 8$ is not sufficient to represent $5/16$ exactly, the result is rounded to either $1/4$ or $3/8$.

To quantify the accuracy of our proposed adder, we compare it to the adder of Fig. 1b with three different SNG configurations: (i) random bit-streams used for the data inputs and an LFSR used for the select input, (ii) random bit-streams for the data inputs and a TFF that toggles every cycle for the select input, and (iii) an LFSR used for the data inputs and a TFF for the select inputs. While the first configuration is more commonly used, we tried two more configurations that provide a slight improvement. However, as seen in Table 2, our proposed adder achieves significantly better accuracy. Once again, the MSEs are calculated by exhaustively testing the adders for every possible input value.

IV. STOCHASTIC-BINARY NEURAL NETWORK DESIGN

We now present our stochastic-binary hybrid design for near-sensor NN computation. Fig. 3 gives an overview of the proposed neural network layer and system design. To evaluate its utility, we will use it to implement the first layer of the LeNet-5 NN topology [19].

*A. Signal Acquisition*

Image sensors capture light intensity and convert it to analog signals, which are converted to digital numbers for processing. In this work, we use parts of a ramp-compare analog-to-digital converter (ADC) to convert the analog signal to the stochastic domain. The conversion circuit shown in Fig. 3 is functionally equivalent to an SNG (Fig. 1c), with some modifications: (i) the inputs are analog, and (ii) a ramp signal is applied to the second input of the comparator rather than a random number generator. Despite becoming heavily auto-correlated, the bit-stream generated by this conversion circuit is still usable for our SC design, because the proposed adder circuits are insensitive to input auto-correlation. Previous work has shown such analog-to-stochastic converters are comparable, in terms of cost and performance to regular ADCs [3][13]. Furthermore, prior work [26] has shown such conversions operate on the order of 100 pJ, which is much lower than the energy consumed by computation (100s of nJ/image). Thus, we do not include the cost of sensor data conversion in our evaluations.

*B. Stochastic Convolutional Neural Network Layer*

The stochastic NN layer consists of 784 stochastic dot-product units shown in Fig. 3 which process the sensor input in parallel. Because there are 32 different first layer kernels, we perform parallel convolutions 32 times per image. The convolution engines perform a basic dot-product operation followed by stochastic-to-binary conversion and an activation function. More precisely, each convolution engine implements:

$$g(\vec{x}, \vec{w}) = sign(\vec{x} \circ \vec{w})$$

where $\vec{x}$ and $\vec{w}$ denote input window and kernel weights, respectively, and $\circ$ denotes the dot-product operation ($\vec{x} \circ \vec{w} = \sum_{i=0}^{k-1} x_i w_i$). The activation function simply outputs the sign of the dot product results and outputs either $-1$, $0$, or $1$. The weight inputs are shared among all convolution engines, so the cost of generating them is amortized across all units.

Since the computation involves negative numbers, the bipolar SC domain $[-1, 1]$ is a natural choice [7]. However, by employing bipolar SC, the decision point of activation functions maps to bit-streams with maximum fluctuation (i.e., unipolar value 0.5). This increases power usage and decreases accuracy. Therefore, we adopt a different approach which uses only unipolar operations by dividing the weights into positive and negative bit-streams $\vec{w}_{pos}$ and $\vec{w}_{neg}$. We then perform two unipolar dot product operations, $g_{pos} = \vec{x} \circ \vec{w}_{pos}$ and $g_{neg} = \vec{x} \circ \vec{w}_{neg}$, followed by two asynchronous counters to convert the results $g_{pos}$ and $g_{neg}$ to the binary domain. Finally, the binary activation function is implemented by a simple comparator. As shown in Fig. 3, the rest of the NN operates in the binary domain.

## V. EXPERIMENTAL RESULTS

This section presents the results of experiments with the proposed SC NN design. We mainly compare our design with a similar all-binary implementation, but when possible we also provide comparisons with existing SC-based NNs.

### A. Experimental Setup

We use the MNIST database [18], a standard machine learning benchmark for handwritten digit recognition, to evaluate accuracy. The benchmark consists of $M = 70,000$ images of handwritten digits (0 to 9); each image uses a 28×28 8-bit greyscale encoding. A subset of 60,000 images are used to train the NN, while the remaining 10,000 images are used to test its accuracy. *Classification accuracy* is defined as the ratio of correctly classified test images to the total number of test images. Then the *misclassification rate* is defined as one minus the classification accuracy. These metrics are often multiplied by 100 and reported as a percentage. All NN training was performed using the TensorFlow framework [1], and the Keras library [10] using a NVIDIA Titan X GPU. For each stochastic design, we built a custom C++ model to evaluate its accuracy.

Previous work on SC NNs [6][16] evaluates NN topologies with only fully connected layers and achieves misclassification rates between 1.95% and 2.41%. On the other hand, our work uses the LeNet-5 topology which has both convolutional and fully connected layers, and achieves misclassification rates around 1%. In practice, the number of convolutional layer kernels and the size of the kernels used in LeNet-5 vary; for our evaluation, we use a variant provided by the Keras library which has the topology shown in Fig. 3.

### B. Accuracy Results and Neural Network Retraining

A key tradeoff in SC is reducing precision to enhance performance. To quantify the impact of reduced precision on classification accuracy, we build separate NN models which execute the first layer of LeNet-5 at different precision levels (2 to 8 bits). We also replace the standard rectified linear activation function with a sign function, which does not impose a significant accuracy loss, but has a much simpler implementation in SC. We do not execute subsequent layers in the stochastic domain since precision losses would compound and require longer bit-streams to achieve accurate results.

For comparison, we evaluate how precision reduction affects the fully binary implementation. Our experiments show that simply quantizing the first layer weights and replacing the activation function with sign detection reduces classification accuracy by several percentage points (up to 6.85% misclassification rate for 4-bit precision). However, by retraining the rest of the NN weights, the NN model is able to recover from the noise introduced by losses in precision and the new activation function (Table 3). Interestingly, we find that we can reduce precision down to 3 or 4 bits and still achieve excellent misclassification rates (below 1%) after retraining. Since the training process is also noisy, the classification accuracy does not always exhibit monotonically decreasing behavior as precision is reduced.

Bit reduction of SC designs exhibits similar accuracy losses, but leads to exponential run time reduction and energy savings. However, stochastic convolutions present unique challenges. SC can be inexact at near-zero input values, and output values are sensitive to errors. Prior work [5] shows that a non-trivial percentage of NN values are near zero, so we use weight scaling and soft thresholding as proposed by Kim et al. [16] to mitigate these errors. Weight scaling normalizes the values of each convolution kernel to use the full dynamic range [−1, 1] while soft thresholding forces a result to zero if it is within some threshold. Finally, we also employ the retraining techniques introduced earlier in the binary domain of the design.

We now compare the resulting classification accuracy using SC with our new adder and multiplier, and the conventional adder and multipliers introduced earlier in Fig. 1 that are used in prior work. Table 3 shows misclassification rates (lower is better) for each design. The results indicate that our new adder and multiplier generally achieve lower misclassification rates than those in prior SC work (up to 2.92% better). We are also able to achieve misclassification rates which are within 0.05% and 0.25% of the binary design for 8-bit and 4-bit precision respectively. Further, the results show that retraining the NN model can compensate for noise introduced by both precision reduction and SC. In particular, for our more accurate adder and multiplication scheme there is less noise that the retraining process must compensate for than the old adder. Note that the benefits of the retraining are only possible because we can operate in the higher precision binary domain. Finally, our results confirm that there is significant opportunity for precision reduction in SC, which translates to exponential reductions in bit-stream lengths and better run times, which we explore next.

Table 3. Misclassification rates for full binary and hybrid stochastic-binary designs, and throughput-normalized power, energy efficiency, and area results for binary and stochastic convolution designs.

|  | Design | 8 Bits | 7 Bits | 6 Bits | 5 Bits | 4 Bits | 3 Bits | 2 Bits |
|---|---|---|---|---|---|---|---|---|
| **Misclassification Rate (%)** | Binary | 0.89% | 0.86% | 0.89% | 0.74% | 0.79% | 0.79% | 1.30% |
|  | Old SC | 2.22% | 3.91% | 1.30% | 1.55% | 1.63% | 2.71% | 4.89% |
|  | This Work | 0.94% | 0.99% | 1.04% | 1.12% | 1.04% | 2.20% | 43.82% |
| **Normalized Power (mW)** | Binary | 40.95 mW | 72.80 mW | 121.52 mW | 204.96 mW | 325.36 mW | 501.76 mW | 683.20 mW |
|  | This Work | 33.17 mW | 33.55 mW | 33.26 mW | 33.01 mW | 33.20 mW | 29.96 mW | 28.35 mW |
| **Energy Efficiency (nJ / frame)** | Binary | 670.92 nJ | 596.38 nJ | 497.74 nJ | 419.76 nJ | 333.17 nJ | 256.90 nJ | 174.90 nJ |
|  | This Work | 543.42 nJ | 274.82 nJ | 136.22 nJ | 67.60 nJ | 34.00 nJ | 15.34 nJ | 7.26 nJ |
| **Area (mm²)** | Binary | 1.313 mm² | 1.094 mm² | 0.891 mm² | 0.710 mm² | 0.543 mm² | 0.391 mm² | 0.255 mm² |
|  | This Work | 1.321 mm² | 1.282 mm² | 1.240 mm² | 1.200 mm² | 1.166 mm² | 1.110 mm² | 1.057 mm² |

## VI. Power, Area, and Energy Evaluation

We synthesize, place-and-route, and measure power using Synopsys Design Compiler, IC Compiler, and PrimeTime for our design; we also use a 65nm TSMC library. For comparison, we evaluate a sliding window convolution engine as our binary baseline design [23]. Activity factors for power measurement are recorded using traces based on MNIST test images and weights from the TensorFlow model.

Table 3 shows the throughput-normalized power, energy efficiency, and design area for both stochastic and binary convolution designs. Power measurements are throughput-normalized relative to the stochastic design. For instance, a binary design operating at 0.25× the throughput and 2× the power relative to a stochastic design would have a throughput-normalized power of 8× relative to the stochastic design. Since run times of stochastic designs decrease exponentially with lower precision, we find that the binary design must operate at exponentially higher frequency and power to match the increase in throughput. Finally, we find the area and energy costs of the SC number generators are higher than a single SC dot product unit, but the cost is shared and amortized over many units.

Since the actual operating frequency will vary across application demands, we contrast the throughput-normalized power between the stochastic and binary designs. Throughput-normalized power is more representative of energy efficiency since it is more agnostic to the differences in frequency and number of parallel units in the design. In terms of energy efficiency, our design breaks even with binary designs at 8-bit precision, and is 9.8× more energy efficient at 4-bit precision. Furthermore, it achieves these gains with better classification accuracy than prior work.

Finally, we see that our stochastic convolution design achieves reasonable area overhead relative to the binary one. The stochastic convolution engine exhibits virtually no change in resource utilization since precision in SC only affects the length of the bit-streams. However, binary designs benefit from linear area reductions since reduced precision narrows the datapath. We find that our design achieves roughly the same area as the binary design at 8-bit precision but is 2× larger than the binary design at 4-bit precision.

## VII. Conclusions

We presented a convolutional NN system which employs a hybrid stochastic-binary design for near-sensor computing. The design employs near-sensor SC using a novel stochastic adder which is significantly more accurate than previous adder designs. Our simulations show that with this adder, the hybrid NN achieves up to 2.92% better accuracy than previous SC designs, and 9.8× better energy efficiency for convolutions over all-binary designs. Finally, we show that retraining the binary domain portion of the NN can compensate for precision losses from SC. As NNs become increasingly commonplace in modern applications, the energy efficiency gains offered by SC will be invaluable for meeting the aggressive power and energy budgets of next generation sensors and embedded devices.

## VIII. acknowledgements

This work was supported in part by the National Science Foundation under Grant CCF-1318091 and Grant CCF-1518703, and generous gifts from Oracle Labs and Microsoft.